\documentclass[]{IEEEtran}
\usepackage{cite}
\usepackage{amsmath,amssymb,amsfonts}
\usepackage{algorithmic}
\usepackage{graphicx}
\usepackage{textcomp}
\usepackage{xcolor}

\ifCLASSOPTIONcompsoc
    \usepackage[caption=false, font=normalsize, labelfont=sf, textfont=sf]{subfig}
\else
\usepackage[caption=false, font=footnotesize]{subfig}
\fi

\usepackage{url}								
\usepackage{siunitx}							
\usepackage{booktabs}							
\usepackage{tabularx}
\newcolumntype{Y}{>{\centering\arraybackslash}X} 

\usepackage{makecell}							
\usepackage{multirow}							
\usepackage[display-foreign=false]{acro}			
\usepackage{gensymb}
\acsetup{single}								

\usepackage{tikz}								

\usepackage{soul}
\usepackage[most]{tcolorbox}

\DeclareAcronym{ls}{
	short = LS,
	long = least-squares
}
\DeclareAcronym{qls}{
	short = QLS,
	long = quadratic least-squares
}
\DeclareAcronym{sinc-ls}{
	short = sinc-LS,
	long = sinc nonlinear least-squares
}
\DeclareAcronym{mf-ls}{
	short = MFLS,
	long = matched filter least-squares
}
\DeclareAcronym{rf}{
	short = RF,
	long = radio frequency
}
\DeclareAcronym{lfm}{
	short = LFM,
	long = linear frequency modulation
}
\DeclareAcronym{prf}{
	short = PRF,
	long = pulse repetition frequency
}
\DeclareAcronym{pri}{
	short = PRI,
	long = pulse repetition interval
}
\DeclareAcronym{fmcw}{
	short = FMCW,
	long = frequency modulated continuous-wave
}
\DeclareAcronym{lfmcw}{
	short = LFMCW,
	long = linear frequency modulated continuous-wave
}
\DeclareAcronym{cw}{
	short = CW,
	long = continuous-wave
}
\DeclareAcronym{dbf}{
	short = DBF,
	long = digital beamforming
}
\DeclareAcronym{sar}{
	short = SAR,
	long = synthetic aperture radar
}
\DeclareAcronym{psr}{
	short = PSR,
	long = point scatterer response
}
\DeclareAcronym{rcs}{
	short = RCS,
	long = radar cross-section
}
\DeclareAcronym{crlb}{
	short = CRLB,
	long = Cramer-Rao lower bound
}
\DeclareAcronym{dof}{
	short = DoF,
	long = degree of freedom
}
\DeclareAcronym{snr}{
	short = SNR,
	long = signal-to-noise ratio
}
\DeclareAcronym{sinr}{
	short = SINR,
	long = signal-to-interference-plus-noise ratio
}
\DeclareAcronym{fft}{
	short = FFT,
	long = fast Fourier transform,
}
\DeclareAcronym{ifft}{
	short = IFFT,
	long = inverse \ac{fft},
}
\DeclareAcronym{ift}{
	short = IFT,
	long = inverse Fourier transform,
}
\DeclareAcronym{rms}{
	short = RMS,
	long = root-mean-square
}
\DeclareAcronym{rmse}{
	short = RMSE,
	long = \ac{rms} error
}
\DeclareAcronym{psd}{
	short = PSD,
	long = power spectral density
}
\DeclareAcronym{rca}{
	short = RCA,
	long = range of closest approach
}
\DeclareAcronym{rda}{
	short = RDA,
	long = Range-Doppler Algorithm
}
\DeclareAcronym{rma}{
	short = RMA,
	long = Range Migration Algorithm
}
\DeclareAcronym{pfa}{
	short = PFA,
	long = Polar Formatting Algorithm
}
\DeclareAcronym{bpa}{
	short = BPA,
	long = Backprojection Algorithm
}
\DeclareAcronym{rvp}{
	short = RVP,
	long = residual video phase
}
\DeclareAcronym{jrc}{
	short = JRC,
	long = joint radar-communications
}
\DeclareAcronym{doa}{
	short = DOA,
	long = direction of arrival
}
\DeclareAcronym{hci}{
	short = HCI,
	long = human-computer interaction
}
\DeclareAcronym{its}{
	short = ITS,
	long = intelligent transportation systems
}
\DeclareAcronym{rtk}{
	short = RTK,
	long = real-time kinematic
}
\DeclareAcronym{eirp}{
	short = EIRP,
	long = effective isotropic radiated power
}
\DeclareAcronym{gnss}{
	short = GNSS,
	long = global navigation satellite system
}
\DeclareAcronym{imu}{
	short = IMU,
	long = inertial measurement unit
}

\DeclareAcronym{ofdm}{
	short = OFDM,
	long = orthogonal frequency division multiplexing
}
\DeclareAcronym{los}{
	short = LoS,
	long = line of sight
}

\DeclareAcronym{pll}{
	short = PLL,
	long = phase-locked loop
}
\DeclareAcronym{vco}{
	short = VCO,
	long = voltage-controlled oscillator
}
\DeclareAcronym{lna}{
	short = LNA,
	long = low-noise amplifier
}
\DeclareAcronym{if}{
	short = IF,
	long = intermediate frequency
}
\DeclareAcronym{cots}{
	short = COTS,
	long = commercial off-the-shelf
}
\DeclareAcronym{adc}{
	short = ADC,
	long = analog to digital converter
}
\DeclareAcronym{dac}{
	short = DAC,
	long = digital to analog converter
}
\DeclareAcronym{lo}{
	short = LO,
	long = local oscillator
}
\DeclareAcronym{pcb}{
	short = PCB,
	long = printed circuit board
}
\DeclareAcronym{mimo}{
	short = MIMO,
	long = multiple-input multiple-output
}
\DeclareAcronym{simo}{
	short = SIMO,
	long = single-input multiple-output
}
\DeclareAcronym{mmic}{
	short = MMIC,
	long = monolithic microwave integrated circuit
}
\DeclareAcronym{daq}{
	short = DAQ,
	long = data acquisition
}
\DeclareAcronym{ic}{
	short = IC,
	long = integrated circuit
}
\DeclareAcronym{pa}{
	short = PA,
	long = power amplifier
}

\DeclareAcronym{ti}{
	short = TI,
	long = Texas Instruments
}
\DeclareAcronym{adi}{
	short = ADI,
	long = Analog Devices
}

\DeclareAcronym{roi}{
	short = ROI,
	long = region of interest,
	long-plural-form = regions of interest
}
\DeclareAcronym{v2x}{
	short = V2X,
	long = vehicle-to-everything
}
\DeclareAcronym{av}{
	short = AV,
	long = automated vehicle
}
\DeclareAcronym{cors}{
	short = CORS,
	long = continuously operating reference station
}
\DeclareAcronym{mdot}{
	short = MDOT,
	long = Michigan Department of Transportation
}
\DeclareAcronym{moco}{
	short = MOCO,
	long = motion compensation
}
\DeclareAcronym{sdr}{
	short = SDR,
	long = software-defined radio
}
\DeclareAcronym{gpio}{
	short = GPIO,
	long = general-purpose input/output
}
\DeclareAcronym{usrp}{
	short = USRP,
	long = Universal Software Radio Peripheral
}
\DeclareAcronym{uhd}{
	short = UHD,
	long = \ac{usrp} Hardware Driver
}
\DeclareAcronym{ntp}{
	short = NTP,
	long = network time protocol
}
\DeclareAcronym{ptp}{
	short = PTP,
	long = precision time protocol
}
\DeclareAcronym{lan}{
	short = LAN,
	long = local area network
}
\DeclareAcronym{wlan}{
	short = WLAN,
	long = wireless \ac{lan}
}
\DeclareAcronym{lut}{
	short = LUT,
	long = lookup table
}
\DeclareAcronym{wsn}{
	short = WSN,
	long = wireless sensor network
}
\DeclareAcronym{mac}{
	short = MAC,
	long = media access control
}
\DeclareAcronym{pps}{
	short = PPS,
	long = pulse-per-second
}
\DeclareAcronym{fom}{
	short = FoM,
	long = figure of merit
}
\DeclareAcronym{uwb}{
	short = UWB,
	long = ultra-wideband
}

\ifcsname hlon\endcsname%
  \newcommand{\hlb}[1]{\textcolor{blue}{#1}}
\else%
  \newcommand{\hlb}[1]{#1}
\fi%

\newtcolorbox{hlbox}[1][]{%
  colback=white,
  float=htb,
  boxsep=0pt,
  #1%
}

\DeclareMathOperator*{\argmax}{argmax} 

\def\BibTeX{{\rm B\kern-.05em{\sc i\kern-.025em b}\kern-.08em
    T\kern-.1667em\lower.7ex\hbox{E}\kern-.125emX}}
\begin{document}

\title{Wireless Picosecond Time Synchronization for Distributed Antenna Arrays}

\author{ Jason M. Merlo,~\IEEEmembership{Graduate~Student~Member,~IEEE,} Serge R. Mghabghab,~\IEEEmembership{Member,~IEEE,}\\and Jeffrey A. Nanzer,~\IEEEmembership{Senior Member,~IEEE}%
 	\thanks{Copyright \copyright 2022 IEEE. Personal use of this material is permitted. However, permission to use this material for any other purposes must be obtained from the IEEE by sending a request to pubs-permissions@ieee.org.}
\thanks{This work was supported under the auspices of the U.S. Department of Energy by Lawrence Livermore National Laboratory under Contract DE-AC52-07NA27344, by the LLNL-LDRD Program under Project No. 22-ER-035, by the Office of Naval Research under grant \#N00014-20-1-2389, and by the National Science Foundation under Grant \#1751655. \textit{(Corresponding author: Jeffrey A. Nanzer)}}
	\thanks{	
	J. M. Merlo and J. A. Nanzer are with the Department of Electrical and Computer Engineering, Michigan State University, East Lansing, MI 48824 USA (email: merlojas@msu.edu, nanzer@msu.edu). S. R. Mghabghab was with the Department of Electrical and Computer Engineering, Michigan State University, East Lansing, MI 48824 USA. He is now with MathWorks, Natick, MA 01760 USA (email: mghabgha@msu.edu).}
}

\maketitle

\begin{abstract}
Distributed antenna arrays have been proposed for many applications ranging from space-based observatories to automated vehicles. Achieving good performance in distributed antenna systems requires stringent synchronization at the wavelength and information level to ensure that the transmitted signals arrive coherently at the target, or that scattered and received signals can be appropriately processed via distributed algorithms. In this paper we address the challenge of high precision time synchronization to align the operations of elements in a distributed antenna array and to overcome time-varying bias between platforms primarily due to oscillator drift. We use a spectrally sparse two-tone waveform, which obtains approximately optimal time estimation accuracy, in a two-way time transfer process. We also describe a technique for determining the true time delay using the ambiguous two-tone matched filter output, and we compare the time synchronization precision of the two-tone waveform with the more common \acf{lfm} waveform. We experimentally demonstrate wireless time synchronization using a single pulse \SI{40}{\mega\hertz} two-tone waveform over a \SI{90}{\centi\meter} \SI{5.8}{\giga\hertz} wireless link in a laboratory setting, obtaining a timing precision of \SI{2.26}{\pico\second}.
\end{abstract}


\acresetall 

\section{Introduction}

\begin{figure}
	\centering
	\includegraphics{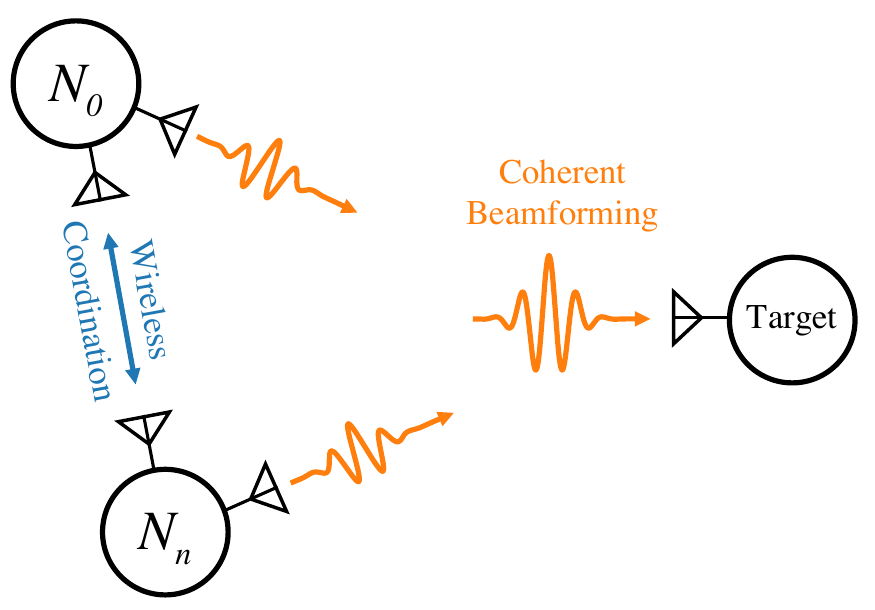}
	\caption{Distributed antenna array schematic. Distributed antenna nodes coordinate wirelessly to align time, frequency, and phase to achieve coherent information summation at the target location.}
	\label{distributed-array}
\end{figure}

\IEEEPARstart{D}{istributed antenna arrays} are rapidly evolving as an essential enabling technology for a variety of novel applications ranging from next generation radio astronomy observatories, small-sat \acf{mimo} communication relays\cite[TX05.2.6, TX08.2.3]{nasa2020taxonomy}, and planetary remote sensing\cite[TA 5.6.7]{nasa2015roadmap}, to collaborative \ac{av} environmental imaging\cite{tagliaferri2021cooperative}. 
Distributed antenna arrays (Fig. \ref{distributed-array}) have a number of benefits over traditional platform-centric approaches.
In conventional single-platform systems, obtaining greater performance requires increasing the aperture size, the power limit or efficiency of the amplifiers, or similar means. However, these approaches are limited by device technologies and platform size, among others, making it increasingly challenging to improve wireless performance.
In a distributed array architecture, many smaller nodes can be used to synthesize the required gain, potentially at a much lower cost than a monolithic array. Additionally, the distributed nature of the system ensures that the array is resilient to node failures or interference, and is furthermore reconfigurable and adaptable, and can thus meet dynamic requirements. In these new distributed aperture applications, however, it is critical that the time, phase, and frequency of the nodes in the array are carefully synchronized to ensure coherent summation of the signal carrier frequencies and alignment of the information envelope at a given target location \cite{nanzer2021distributed}.  While there have been many significant advances in the areas of wireless time, frequency, and phase coordination between nodes in distributed arrays, there are still significant advances required in each of these areas to enable the continuous high accuracy coordination required to provide coherent operation at millimeter-wave frequencies and multi-gigahertz information bandwidths. 
In particular, for modulated waveforms with wide bandwidth, accurate time alignment, i.e., clock synchronization, is critical to ensure high coherent gain at the target location~\cite{nanzer2017open}.

While optical means have been used for disciplining remote oscillators and clock alignment wirelessly to femtosecond and sub-femtosecond levels by exploiting the large available bandwidth, typically in the terahertz \cite{giorgetta2013optical, sinclair2016synchronization, sinclair2019femtosecond}, the pointing and tracking tolerance for moving targets is very tight making it significantly more difficult to implement for dynamic links; in addition, the size, weight, and cost of free-space optical systems is often higher than for microwave and millimeter-wave systems. Because of these limitations to optical links, it is of interest to develop microwave and millimeter-wave wireless time synchronization techniques. There have been many prior works focusing on microwave and millimeter-wave time synchronization of \acp{wsn}\cite{sichitiu2003simple,ganeriwal2003timing,hill2002wireless}, however, these have focused primarily on achieving synchronization at the protocol level, in some cases with hardware timestamping at the \ac{mac} layer \cite{hill2002wireless} to reduce timing uncertainties, but were primarily motivated by synchronization of higher-level protocols and data logging where coordination at the microsecond-level was sufficient. The more stringent requirements of distributed beamforming and high bandwidth communications necessitates improvements of several orders of magnitude over previous \ac{wsn} techniques. 
In recent years, this has been approached via increased signal bandwidth.
\hlb{One recent approach implements a wireless White Rabbit-based protocol using a \textit{V}-band carrier with a \SI{1.6}{\giga\hertz} bandwidth to achieve a precision of $<\,$\SI{2}{\pico\second} over a $\sim$\SI{500}{\meter} \ac{los} link~\cite{gilligan2020white}.  Another approach using \SI{50}{\mega\hertz} \ac{lfm} waveforms recently achieved synchronization precision of \SI{11.3}{\pico\second} with a \ac{snr} of $31.2$\,dB in an outdoor \ac{los} environment}~\cite{prager2020wireless}. 
A third approach using an ``enhanced timestamping'' cross-correlation approach on top of the IEEE 802.11n \ac{wlan} standard with a carrier frequency of \SI{2.412}{\giga\hertz} and \SI{20}{\mega\hertz} bandwidth achieved a timing precision of approximately \SI{650}{\pico\second} \cite{seijo2020enhanced}.

In this paper we 
demonstrate a new technique for the high precision estimation of time delay in a two-way time transfer system for distributed array applications.  By utilizing a spectrally sparse two-tone waveform it is shown that the mean-squared-bandwidth of the time delay estimation waveform may be maximized, which yields the maximum theoretical accuracy for time delay estimation. Using this waveform, we experimentally demonstrate a wireless time synchronization precision of $<$\SI{2.5}{\pico\second} using single pulse time estimation with a waveform bandwidth of \SI{40}{\mega\hertz} in commercial \acp{sdr}. 
\hlb{This work is the first to demonstrate the use of a spectrally sparse two-tone waveform in a fully wireless coordination approach. In prior work we briefly introduced a two-way time synchronization approach using a two-tone waveform~\cite{merlo2022aps}, however that work required the use of a cabled frequency reference. Here we combine wireless time transfer with wireless frequency locking to provide a fully wireless approach and demonstrate the ability to obtain picosecond-level time synchronization between nodes.
We provide a significantly more detailed description of the system implementation, discussing the time-delay estimation processes used, their challenges, and techniques to mitigate the challenges.
We also describe the \ac{crlb} for time delay estimation and how to maximize the accuracy of the time delay waveform to achieve the theoretical maximum accuracy for a given signal bandwidth and \ac{snr}. 
Finally, we present fully cabled and fully wireless time and frequency synchronization experiments and compare their relative performance to the \ac{crlb}. We furthermore evaluate the long-term beamforming channel bias measurements to demonstrate the long-term system synchronization bounds due to the current hardware limitations.}

The rest of the paper is organized as follows. In Section \ref{sec:two-way-time-synchronization} we introduce the system time model and two-way time synchronization process, then proceed with the derivation of the \ac{crlb} for time delay estimation of the conventional \ac{lfm} and two-tone waveforms, and finally, we discuss our two-step delay estimation process. In Section \ref{sec:frequency-synchronization} we discuss the details of frequency synchronization in distributed arrays and the approach used in this paper. Finally, in Section \ref{sec:experiments} we discuss the system hardware configuration and the results of the time-transfer precision experiments 
\hlb{for two-tone waveforms over a range of \ac{snr} levels for three cases: fully cabled, wireless time-transfer with cabled frequency transfer, and fully wireless time-frequency transfer scenarios. Finally, we provide a comparison of other wireless microwave and millimeter-wave time transfer methods as a benchmark for the proposed technique using spectrally sparse waveforms.}
\section{Distributed Antenna Array Two-way Time Synchronization}
\label{sec:two-way-time-synchronization}

Generally, two types of techniques are commonly employed to synchronize distributed clocks: one-way methods, and two-way methods.  The most common technique for wireless time transfer is one-way as it is employed by many of the \ac{gnss} constellations in orbit today for time distribution where the \ac{gnss} satellite acts as a ``primary'' clock source and all the receiver nodes synchronize their clocks to the primary source after solving for the propagation delay of the signal based on the ephemeris provided by each satellite and the known position of the receiver \cite{levine2008review}. The difficulty of this process is that either the receiver and transmitter's positions must both be known, or multiple sources with known positions are required to solve for the propagation delay.  An alternative approach is two-way time synchronization, which inherently solves for both the time of flight and clock offset, assuming a quasi-static channel during the synchronization epoch; the two-way time transfer technique has been used for satellite time transfer for many decades to synchronize satellites to ground clocks and with other satellites \cite{cooper1979review,hanson1989fundamentals,kirchner1991two-way}. A more recent protocol which uses two-way time transfer is \acf{ptp} which achieves timing precision on the order of \SI{1}{\micro\second}; \ac{ptp} is also the foundation for the White Rabbit protocol, a popular industrial synchronous Ethernet protocol which acts as a refinement on the \ac{ptp} estimation by using the carrier phase of Ethernet over fiber to determine residual time delay with high precision, typically on the order of \SI{10}{\pico\second} \cite{serrano2013white}. However, the White Rabbit protocol is designed to work over fiber and thus cannot be used on its own to coordinate wirelessly. 

\begin{figure}
	\centering
	\includegraphics[width=\columnwidth]{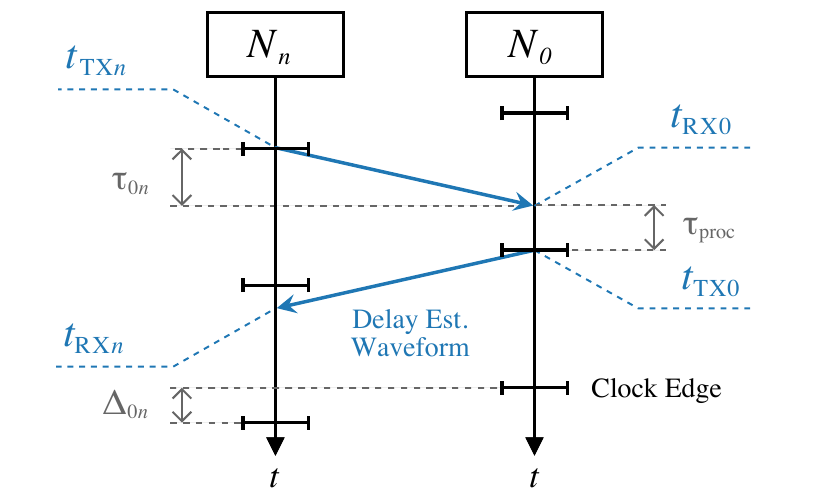}
	\caption{Two-way time transfer timing diagram. Node $N_n$ initiates time transfer with the primary node, $N_0$; a delay estimation waveform is transmitted from node $N_n$ to $N_0$ and back with timestamps saved at each transmission and reception. From the four timestamps, a time offset and inter-node distance can be computed using \eqref{time-delta} and \eqref{time-delay} respectively, assuming that the channel was quasi-static over the synchronization epoch (i.e., between $t_{\mathrm{TX}_n}$ and $t_{\mathrm{RX}_n}$).}
	\label{time-transfer-schematic}
\end{figure}

\subsection{System Model}
In general, a distributed array system can be modeled as a set of $N$ nodes, each of which has a local clock which is mapped to the true global time $t$ by a function 
\begin{equation}
	T_{n}(t)=t+\epsilon_n(t)	
\end{equation}
where $\epsilon_n(t)$ is a time-varying bias which consists of a time-varying frequency offset random walk and a noise term consisting of thermal noise, shot noise, flicker noise, plasma noise, and quantum noise, among other sources, depending on device technology \cite[chapter 10.1]{pozar2005microwave}.  \hlb{In this work, we assume any time-varying frequency offset and the time offset $\delta_n(t)$ are quasi-static over the synchronization epoch; in the experiments discussed in this paper, the systems are furthermore syntonized (synchronized in frequency) and thus, over a long term, may be treated as a constant bias plus a noise term}
\begin{equation}
	\label{error}
	\hlb{\epsilon_n(t)=\delta_n(t)+\nu_n(t)}
\end{equation}
\hlb{where $\delta_n(t)$ is the quasi-static time offset of node $n$ relative to the global time during the synchronization epoch and $\nu_n(t)$ is the noise at node $n$ at time $t$. The focus of this work is estimating and correcting for quasi-static bias term $\delta_n(t)$ which consists of static and dynamic components; dynamic components include frequency offset between platforms as well as time-varying internal delays caused by thermal expansion and nonlinear components whose propagation delay varies with environmental parameters; static components include constant system delays due to trace and cable lengths internal to the system which can be calibrated out.} To simplify the model, it is assumed that node $0$ is the true global time, thus the bias of node zero is $\delta_0=0$, and the goal is to find \mbox{$\Delta_{0n}=\delta_0-\delta_n$}.

\subsection{Two-way Time Transfer}
\label{two-way-time-syncrhonization}

In a two-way time transfer system, synchronization is achieved by sending a time delay estimation waveform between two nodes in both directions, schematically pictured in Fig. \ref{time-transfer-schematic}.  Assuming the link is quasi-static during the synchronization epoch, the offset between the local clock at node $n$ and node $0$ can be deduced by
\begin{equation}
	\label{time-delta}
	\Delta_{0n}=\frac{\left(t_{\mathrm{RX}0}-t_{\mathrm{TX}n}\right)-\left(t_{\mathrm{RX}n}-t_{\mathrm{TX}0}\right)}{2}
\end{equation}
where $t_{\mathrm{TX}n}$ and $t_{\mathrm{RX}n}$ are the times of transmission and reception at node $n$ respectively.  Once this offset is estimated it may be added to the local clock at node $n$ to compensate for the accumulated bias.  Note that $\tau_\mathrm{proc}$, the processing time between the initial pulse reception at node $n$ and its response, is arbitrary and does not affect the ability to determine the time offset so long as the assumption that the clock bias is quasi-static over the synchronization epoch is valid. Further clock characterization could be inferred by taking statistics over long-term bias correction to determine a constant drift between platforms which could be tracked using techniques such as Kalman filtering to improve time stability between synchronization exchanges. If the link is symmetric, the propagation delay can also be deduced simply by
\begin{equation}
	\label{time-delay}
	\tau_{0n}=\frac{\left(t_{\mathrm{RX}0}-t_{\mathrm{TX}n}\right)+\left(t_{\mathrm{RX}n}-t_{\mathrm{TX}0}\right)}{2}.
\end{equation}

\subsection{Theoretical Bounds on Time Delay Accuracy}
\label{accuracy-bounds}
The theoretical limit on the ability to accurately estimate the time delays in the above is dependent on the \ac{snr} and the waveform characteristics. The limit is given by the \ac{crlb}, which defines the variance on the estimate of the delay as~\cite[Chapter 7.2]{richards2014fundamentals},  \cite{nanzer2017accuracy},
\begin{equation}
	\label{crlb}
	\mathrm{var}(\hat{\tau}-\tau)\ge\frac{N_0}{2\zeta_f^2 E_s} \\
\end{equation}
where $\zeta_f^2$ is the mean-squared bandwidth (the second moment of the spectrum of the signal), $E_s$ is the signal energy, and $N_0$ is the noise \ac{psd}, where  
\begin{equation}
	\label{tbp}
	\frac{E_s}{N_0} = \tau_\mathrm{p}\cdot\mathrm{SNR}\cdot\mathrm{NBW}
\end{equation}
where $\tau_\mathrm{p}$ is the pulse duration, \ac{snr} is the pre-processed \ac{snr}, and \textrm{NBW} is the noise bandwidth of the system. It is clear from \eqref{crlb} and \eqref{tbp} that the variance of the time delay estimate is inversely proportional to the \ac{snr} and mean-squared bandwidth of the waveform used.  Thus, by increasing coherent integration time and transmission power, and mitigating channel and system noise, the variance of the time delay estimate may be reduced.  However, of greater interest is designing a waveform which maximizes the mean-squared bandwidth to obtain the highest accuracy theoretically possible for any given \ac{snr} level. In \cite{nanzer2017accuracy} it is shown that the mean-squared bandwidth for a waveform can be represented by
\begin{equation}
	\label{mean-squre-bandwidth}
	\zeta_f^2=\int_{-\infty}^{\infty} \left(2\pi f\right)^2 \left|G(f)\right|^2 df
\end{equation}
where $G(f)$ is the \ac{psd} of the signal. From \eqref{mean-squre-bandwidth} it can be shown that concentrating the \ac{psd} of the waveform to the edges of the spectrum in a given bandwidth, yielding a two-tone waveform, maximizes the mean-squared bandwidth of the waveform, thus minimizing the estimation variance \eqref{crlb}. Computing the mean-squared bandwidth for a fully filled bandwidth waveform, such as the \ac{lfm} yields
\begin{equation}
	\label{msb-lfm}
	\zeta_{f\textrm{(LFM)}}^2=\frac{\left(\pi\cdot\mathrm{BW}\right)^2}{3}
\end{equation}
where $\mathrm{BW}$ is the maximum extent of the waveform bandwidth, whereas a two-tone waveform with its energy located at the edges of its bandwidth yields
\begin{equation}
	\label{msb-two-tone}
	\zeta_{f\textrm{(two-tone)}}^2=\left(\pi\cdot\mathrm{BW}\right)^2
\end{equation}
an improvement by a factor of three. A waveform consisting of two tones at the edges of the spectrum is the optimal form of the time delay estimation waveform. This finding not only yields improved delay estimation but may also reduce system requirements as very large bandwidths can be synthesized using only two instantaneously narrow-band transmitters as opposed to a single wide-band transmitter which are often very difficult to design and calibrate to ensure a uniform power response across the entire operating bandwidth. This channelized approach to two-tone high accuracy delay estimation for range measurement was demonstrated experimentally in \cite{schlegel2019microwave}.


\subsection{Time Delay Estimation and Refinement Process}
\label{sec:delay-estimation}
To estimate the time delay of the received two-tone waveform, a matched filter is used which maximizes the signal energy at the output of the filter at the time delay of the start of the received waveform. For a discretely sampled waveform the matched filter output is
\begin{align}
	\begin{aligned}
	s_\mathrm{MF}[n] &= s_\mathrm{RX}[n]\circledast s_\mathrm{TX}^*[-n]\\
	& = \mathcal{F}^{-1}\left\{{S_\mathrm{RX}S_\mathrm{TX}^*}\right\}
	\end{aligned}
\end{align}
where $s_\mathrm{TX}$ is the ideal transmitted waveform, $s_\mathrm{RX}$ is the received waveform, and $( \cdot )^*$ is the complex conjugate \cite{richards2010principles,mghabghab2021microwave}. While a continuous-time matched filter maximizes the output power at the true time delay of the received waveform, the discrete-time matched filter produces a peak at the sample bin most closely corresponding to the true time delay of the received signal; this is an issue for high accuracy time delay estimation as it implies the estimation accuracy is limited by the sample rate of the digitizer. This may be overcome by increasing the sample rate, however, for high-accuracy requirements, this rapidly becomes prohibitively expensive using current hardware as sample rates exceed multiple gigahertz.
\begin{figure}
	\includegraphics[width=\columnwidth]{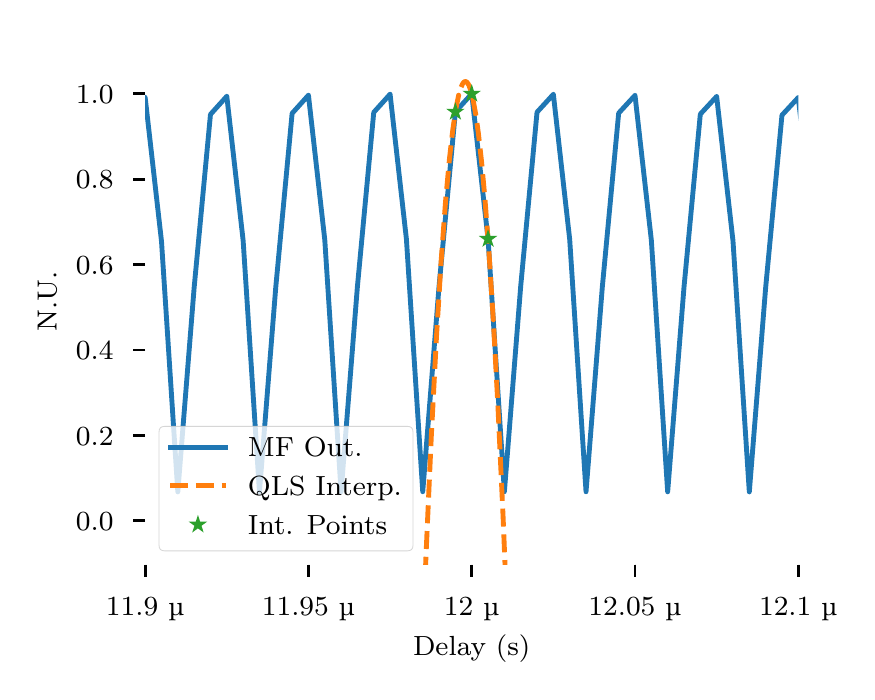}
	\caption{Peak region of the discrete matched filter output for an ideal two-tone waveform (blue line). \Acf{qls} refinement (orange dashed line) is used to interpolate between sample points using the matched filter peak and two adjacent points (green dots) to mitigate discretization error.}
	\label{mf-output}
\end{figure}
In contrast, a two-stage estimation can be employed where the coarse delay is estimated at the resolution of the data converter and then refined via processing to more accurately estimate the true time delay. One simple technique is \acf{qls} interpolation in which a parabola is fitted to the peak of the matched filter and the two adjacent points; the time delay at the peak of the parabola is then regarded as the true time delay, graphically depicted in Fig. \ref{mf-output}. The peak of a parabola formed by the peak of the discrete matched filter and its two adjacent sample points may be easily found in constant time complexity by \cite{moddemeijer1991sampled}, \cite[Chapter 7.2]{richards2014fundamentals}
\begin{align}
	n_{\mathrm{max}} &= \argmax_n\left\{s_{\mathrm{MF}}[n]\right\}\\
	\hat{\tau} &= \frac{T_{\mathrm{s}}}{2} \frac{s_{\mathrm{MF}}[n_{\mathrm{max}} - 1]-s_{\mathrm{MF}}[n_{\mathrm{max}} + 1]}{s_{\mathrm{MF}}[n_{\mathrm{max}}-1] - 2 s_{\mathrm{MF}}[n_{\mathrm{max}}] + s_{\mathrm{MF}}[n_{\mathrm{max}}+1]}
\end{align}
where $T_{\mathrm{s}}$ is the sampling interval. 
\begin{figure}
	\includegraphics[width=\columnwidth]{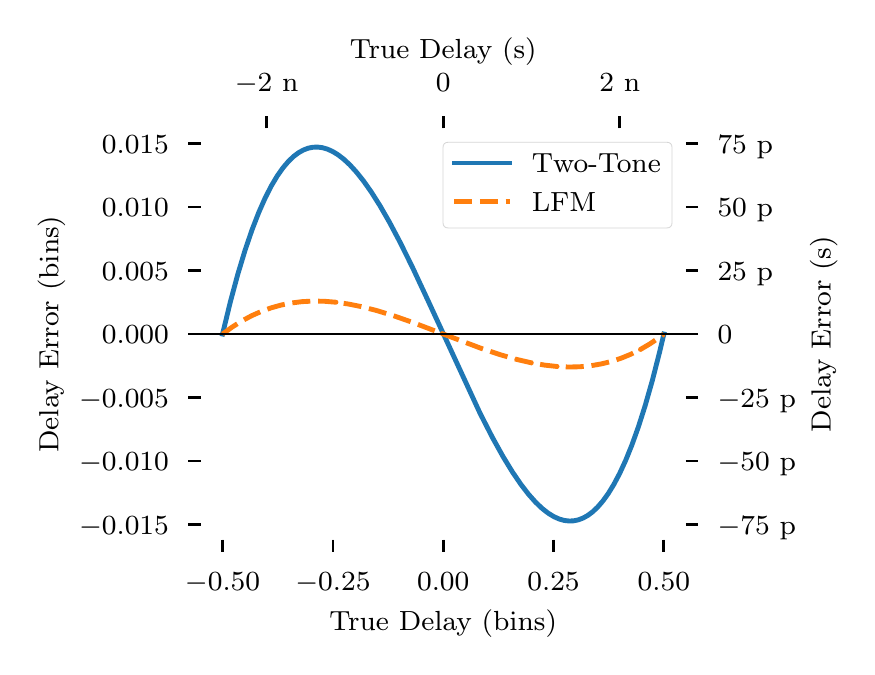}
	\caption{Waveform and sample-rate dependent residual bias after \Acl{qls} interpolation shown for \SI{40}{\mega\hertz} two-tone and \ac{lfm} waveforms. Given the cooperative nature of two-way time transfer, a \acf{lut} based on these curves may be used to correct for this bias by subtracting the bias at the given fractional true delay bin.}
	\label{qls-bias}
\end{figure}

\ac{qls} can greatly reduce the discretization errors introduced by the sample rate. However, if the underlying matched filter does not perfectly match a parabola, a residual delay-dependent bias will manifest in the inter-sample period $1/T_{\mathrm{s}}$ that is inversely proportional to the \ac{qls} oversampling ratio, i.e., the factor by which the sampler exceeds the Nyquist frequency of the time delay waveform being sampled.  The bias can be seen in Fig. \ref{qls-bias} for the \ac{lfm} and two-tone waveforms.  It is also important to note that the shape of the bias is not sinusoidal and is, furthermore, dependent on the waveform parameters. While the bias was present in both \ac{lfm} and two-tone waveforms, the two-tone waveforms were found to have larger biases using this technique: e.g., for a two-tone waveform sampled at \SI{200}{\mega Sa/s} with a tone separation of \SI{40}{\mega\hertz}, a peak bias of $\sim$\SI{73}{\pico\second} is expected, whereas an \ac{lfm} of equal bandwidth exhibits a peak bias of only $\sim$\SI{13}{\pico\second}, shown in Fig. \ref{qls-bias}. While these are large biases when working towards sub-picosecond levels of precision, they are predictable if the waveform parameters and sample rates are known a priori and can be easily corrected via \acf{lut}.  By precomputing the expected biases at each fractional delay bin and storing the results in a \ac{lut}, the bias may be corrected for at runtime reducing the overall bias due to the estimator to arbitrarily low levels. \hlb{Alternative peak interpolation techniques which more closely match the transmitted waveform include \acf{sinc-ls} and \acf{mf-ls} \cite{mghabhab2022microwave}. \Ac{sinc-ls} more closely approximates the shape of the output of the matched filter for two-tone signals, but is an iterative approach, and thus requires longer computation time than \ac{qls}. \Ac{mf-ls} matches the shape of the output of the matched filter exactly, but also requires the matched filter to be computed iteratively to optimize the fit of the estimated time delay proposal with the received signal which requires significantly longer computation time than \ac{qls}. Due to the simplicity of implementation, low computational complexity, and relatively high accuracy achieved using \ac{qls} with a lookup table for bias correction, the \ac{qls} technique was chosen for use in these experiments.}

\hlb{It should also be noted that while the systems are syntonized, the frequency jitter, captured by $\nu_n(t)$ in \eqref{error}, will still impact the sampling uniformity of the \ac{adc} and \ac{dac}, and thus cause distortion in the respective transmitted and received signals due to sampling nonuniformities; this will have the effect of reducing the accuracy of the matched filter due to distortion of the transmitted and sampled received waveforms resulting in a mismatch between the ideal and sampled waveforms.}

\section{Frequency Synchronization}
\label{sec:frequency-synchronization}
Frequency synchronization, or syntonization, is the process of making the clocks on all platforms resonate with the same period; this is essential for two reasons: 1) to ensure that the signals sum coherently at the destination, and 2) to ensure the transmitted and received waveforms are sampled with the same period to ensure the matched filter correctly estimates the time delay of the received waveform for the time transfer and ranging estimation. 
If time synchronization is sufficiently accurate (i.e., a small fraction of the oscillator period) and is implemented with a sufficiently fast periodicity to minimize oscillator drift, the frequencies on each node can theoretically be synchronized directly by aligning the phases of the oscillators. However, not all systems have the ability to directly adjust the oscillator phase, particularly if time synchronization is added to existing legacy systems or commercial hardware. In these cases, frequency synchronization is also necessary, and may allow for a relatively infrequent time synchronization interval.

There are many ways to accomplish wireless frequency synchronization which broadly fall into three categories: closed-loop, open-loop centralized, and open-loop decentralized. In a closed-loop topology the distributed nodes utilize feedback from a cooperative target which transmits back information used to tune the distributed nodes to the proper transmit frequency \cite{seo2008feedback,bidgare2012implementation}. While this can be useful in communication systems, for targets with passive receivers or radar applications, nodes cannot rely on feedback from a target and must implement open-loop topologies.  In an open-loop centralized architecture, a single primary node is utilized as the ``leader'' which generates the frequency reference for all other nodes to syntonize to \cite{barriac2004distributed,brown2005method,mghabghab2021open-loop}. This approach enables remote observation and communications with passive targets but has a single point of failure at the primary node as well as inherently has an array size limit due to increasing path loss between the primary node and followers as the array size grows.  Finally, the open-loop distributed architecture consists of many nodes which all perform a frequency consensus averaging operation wherein all nodes in the array attempt to estimate the frequency of all other adjacent nodes and adjust their own frequency to the average of the estimates \cite{ouassal2020decentralized, ouassal2021decentralized, rashid2022frequency}.  This approach is the most robust to interference as well as avoids the single point of failure and scaling difficulties of the open-loop centralized architecture, however it is also the most difficult to implement due to the necessity of separately estimating and tracking the frequencies from multiple nodes and performing online adjustment of the local carrier frequency which typically requires a software-based implementation.

\begin{figure} 
	\centering
	\includegraphics{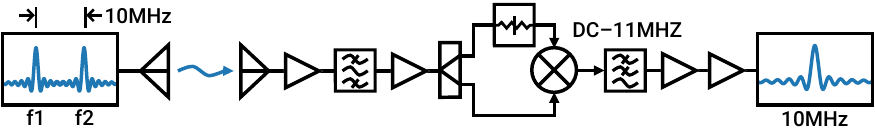}
	\caption{Wireless frequency transfer circuit schematic.  A two-tone waveform at carrier frequency is transmitted with a tone separation of \SI{10}{\mega\hertz}; the two-tone is received, amplified and filtered, then split and self-mixed. The resultant signal consists of a \SI{10}{\mega\hertz} tone with other tones near $2f_{0\mathrm{f}}$ which are easily filtered by a lowpass filter.  The \SI{10}{\mega\hertz} tone is finally amplified by a clock buffer to produce a \SI{10}{\mega\hertz} square wave for frequency reference to the \ac{sdr}.}
	\label{frequency-locking} 
\end{figure}

In this paper, we implement an open-loop centralized approach due to its balance of being able to perform radar and passive target communication operations as well as being relatively simple to implement in a hardware circuit. A spectrally sparse technique using a self-mixing receiver is utilized \cite{mghabghab2020frequency, abari2015airshare} to provide improved robustness compared to single-tone frequency transfer techniques, which are more susceptible to external interference. The principle of operation is shown in Fig.~\ref{frequency-locking}. A two-tone waveform with a tone separation of $\beta_\mathrm{f}=\SI{10}{\mega\hertz}$ is generated at an arbitrary carrier frequency $f_{0\mathrm{f}}$, the tones are received at the self-mixing circuit where out-of-band noise is filtered by a bandpass filter and the signal is amplified and mixed with itself.  This generates tones at the sum and difference of the original received tones resulting in a \SI{10}{\mega\hertz} tone as well as other tones around $2f_{0\mathrm{f}}$, the latter of which are easily removed by a lowpass filter. Finally, the \SI{10}{\mega\hertz} tone is converted to a square wave via a clock amplifier to provide optimal performance for the frequency reference on the \ac{sdr}. \hlb{It should be noted that the tone separation $\beta_\mathrm{t}$ is chosen to be \SI{10}{\mega\hertz} due to the requirement of at \SI{10}{\mega\hertz} reference input on the \acp{sdr}, however this may be any arbitrary value that is advantageous for the device requiring a frequency reference.}

\hlb{Finally, while in these experiments we use separate \ac{rf} bands for the time and frequency transfer waveforms, there is no specific requirement for the waveforms to exist in any given band, since the performance is strictly bandwidth-dependent. Furthermore, it has previously been shown that these two functions can coexist in the same band: in \cite{ellision2020combined} a three-tone waveform was used to accomplish frequency transfer and ranging using a narrow tone separation of \SI{10}{\mega\hertz} for frequency transfer, and a wide tone separation of \SI{200}{\mega\hertz} for range (time delay) estimation.  This could be employed for time-frequency transfer as well by simply pulsing the time transfer tone while keeping the two frequency transfer tones continuous-wave, however, it was chosen to use separate \ac{rf} bands for this experiment for simplicity.}

\section{High Precision Time Transfer Experiments}
\label{sec:experiments}
\subsection{Experimental Configuration}
\begin{figure*}
	\centering
	\includegraphics[width=0.9\textwidth]{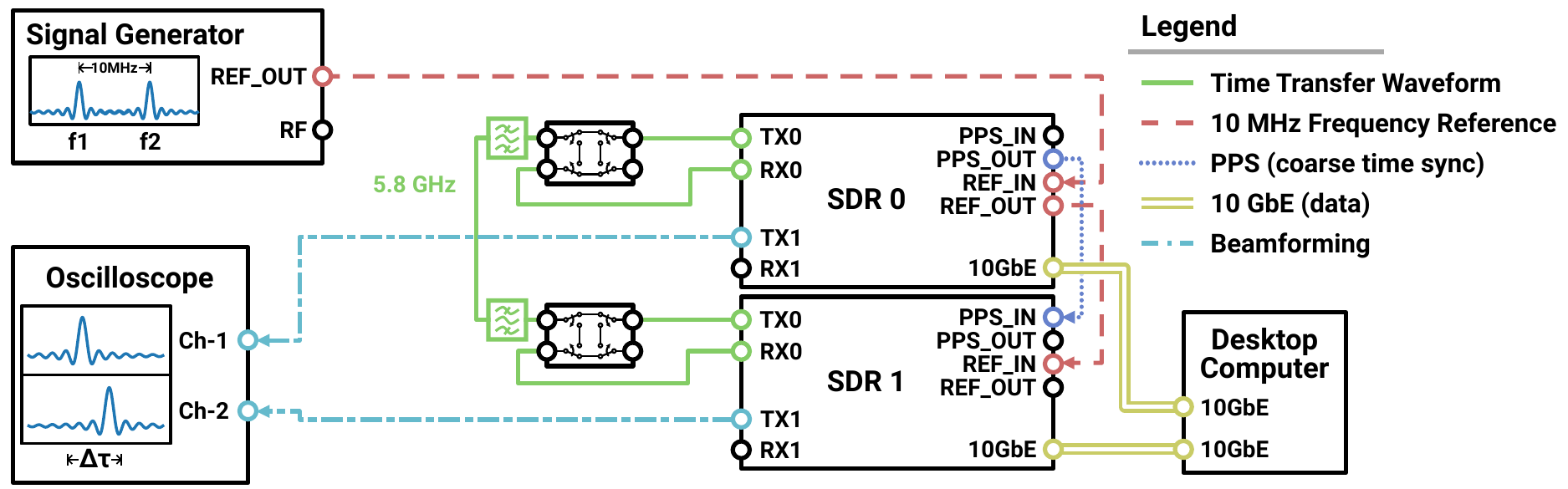}
	\caption{\hlb{Fully cabled time-frequency transfer system schematic. The signal generator is used as the primary frequency reference for \ac{sdr} 0 (primary \ac{sdr}), which provides the frequency reference for \ac{sdr} 1 (secondary \ac{sdr}). The oscilloscope is used to sample and digitize the beamforming waveforms to determine beamforming accuracy while performing time transfer.  Control transfer switches were used to provide high isolation between transmit and receive paths during time-domain multiplexing operation.  Both \acp{sdr} were controlled by a single desktop computer using GNU Radio during operation. A \acf{pps} signal was used on device startup for an initial, coarse time alignment to ensure transmit and receive windows on each \ac{sdr} overlap for the fine time alignment process to proceed. Unused ports on switches were terminated with matched loads.}}
	\label{schematic-cabled}
\end{figure*}
\begin{figure*}
	\centering
	\includegraphics[width=0.9\textwidth]{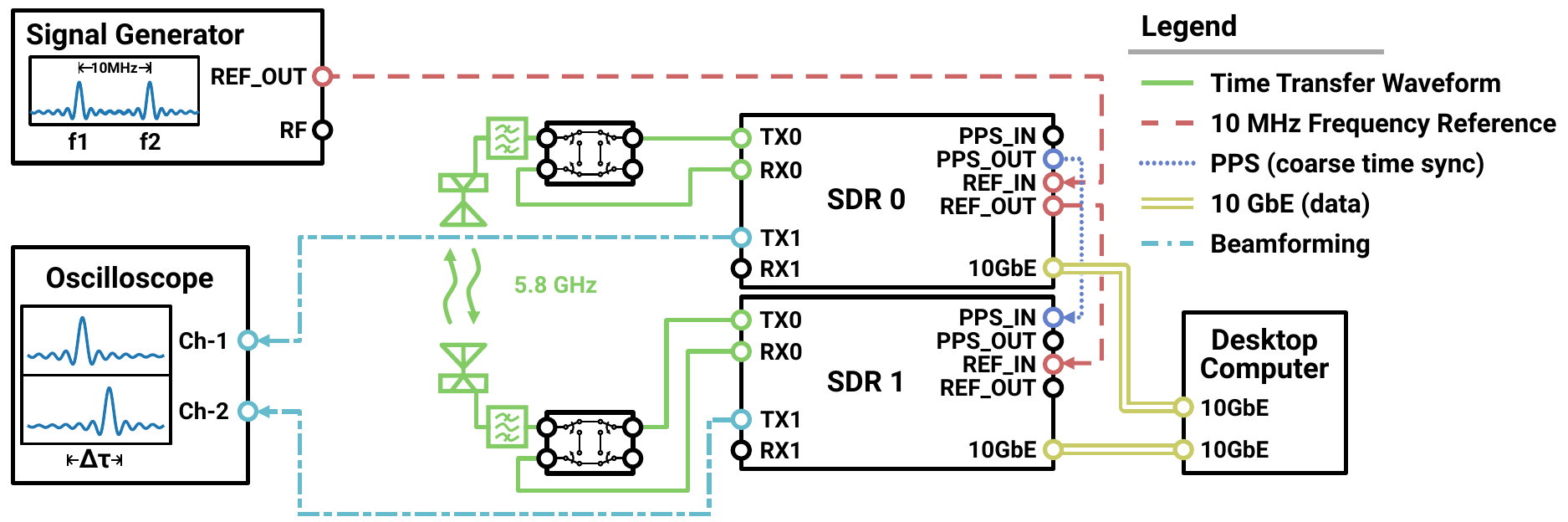}
	\caption{\hlb{Wireless time transfer, cabled frequency syntonization system schematic. The experiment was configured similarly to the fully cabled experiment with the exception time transfer was performed over a wireless link. Unused ports on switches and splitters were terminated with matched loads.}}
	\label{schematic-wireless-tt}
\end{figure*}
\begin{figure*}
	\centering
	\includegraphics[width=0.9\textwidth]{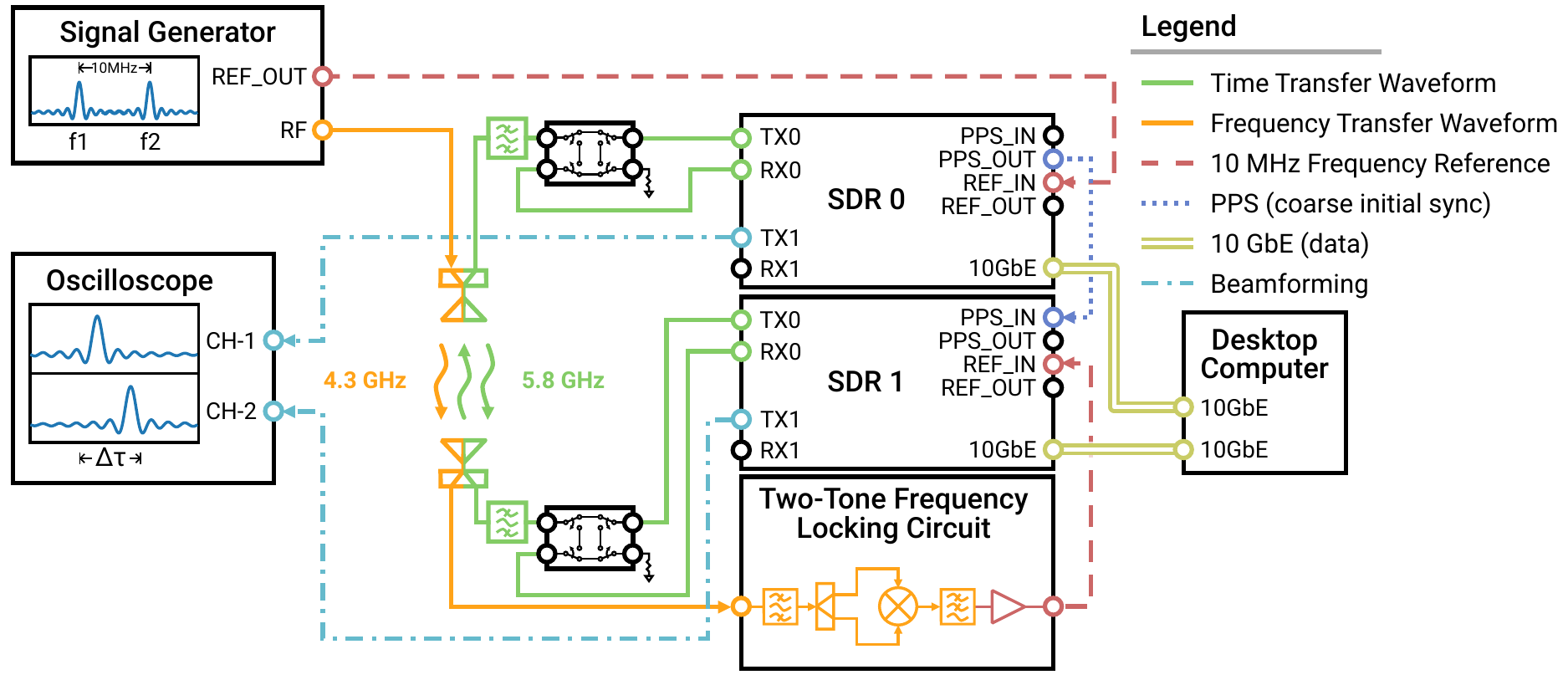}
	\caption{\hlb{Fully wireless time-frequency transfer system schematic. The experiment was configured similarly to the wireless time transfer experiment, however The signal generator was used the two-tone generator for the self-mixing frequency locking circuit, used as the frequency reference for \ac{sdr} 1 (secondary \ac{sdr}). Unused ports on switches and splitters were terminated with matched loads.}}
	\label{schematic}
\end{figure*}
\begin{table}[tb]
	\ifcsname hlon\endcsname%
		\color{blue}%
  	\fi%
	\caption{\hlb{Experiment Parameters}}
	\label{tab:experiment-parameters}
	\begin{center}
  	\begin{tabularx}{\columnwidth}{p{0.43\linewidth}YY}
	
	\toprule[1pt]
	\textbf{Time Transfer Waveform} \\
	\midrule
	\midrule
	Parameter & Symbol & Value \\
	\midrule
	Waveform Type &  & \mbox{Pulsed\;Two-Tone} \\
	Carrier Frequency & $f_{0\mathrm{t}}$ & \SI{5.8}{\giga\hertz} \\
	Tone Separation & $\beta_\mathrm{t}$ & \SI{40}{\mega\hertz} \\
	Rise/Fall Time & & \SI{50}{\nano\second} \\
	Pulse Duration & $\tau_\mathrm{p}$ & \SI{10.0}{\micro\second} \\
	Synchronization Epoch Duration &  & \SI{50.01}{\milli\second} \\
	Resynchronization Interval &  & \SI{100.0}{\milli\second} \\
	Rx Sample Rate & $f^\mathrm{Rx}_\mathrm{s}$ & \SI{200}{\mega Sa/s} \\
	Tx Sample Rate & $f^\mathrm{Tx}_\mathrm{s}$ & \SI{400}{\mega Sa/s}* \\
	\midrule[1pt]
	\textbf{Frequency Transfer Waveform}\\
	\midrule
	\midrule
	Parameter & Symbol & Value \\
	\midrule
	Waveform Type &  & \acs{cw} Two-Tone \\
	Carrier Frequency & $f_{0\mathrm{f}}$ & \SI{4.3}{\giga\hertz} \\
	Tone Separation & $\beta_\mathrm{f}$ & \SI{10}{\mega\hertz} \\
	\midrule[1pt]
	\textbf{Beamforming Waveform} \\
	\midrule
	\midrule
	Parameter & Symbol & Value \\
	\midrule
	Waveform Type &  & \mbox{Pulsed\;Two-Tone} \\
	Carrier Frequency & $f_{0\mathrm{t}}$ & \SI{1.2}{\giga\hertz} \\
	Tone Separation & $\beta_\mathrm{b}$ & \SI{50}{\mega\hertz} \\
	Rise/Fall Time & & \SI{50}{\nano\second} \\
	Pulse Duration & $\tau_\mathrm{p}$ & \SI{10.0}{\micro\second} \\
	Tx Sample Rate & $f^\mathrm{Tx}_\mathrm{s}$ & \SI{400}{\mega Sa/s}* \\ 
	Rx Sample Rate & $f^\mathrm{osc}_\mathrm{s}$ & \SI{20}{\giga Sa/s} \\
	\midrule[1pt]
	\textbf{Antenna Parameters}\\
	\midrule
	\midrule
	Parameter & Symbol & Value \\
	\midrule
	Gain &  & $8$\,dBi\\
	Bandwidth &  & \SI{2.3}{}--\SI{6.5}{\giga\hertz}\\
	Separation (radome-to-radome) & & \SI{90}{\centi\meter} \\
	\bottomrule[1pt]	
	\end{tabularx}
	\end{center}
	\footnotesize{\hlb{* Digitally upsampled from \SI{200}{\mega Sa/s} to \SI{400}{\mega Sa/s} on device}}
\end{table}
\hlb{The time transfer experiments consisted of three configurations:
\begin{itemize}
	\item[1)] fully cabled time-frequency transfer (Fig. \ref{schematic-cabled}),
	\item[2)] wireless time transfer with cabled frequency syntonization (Fig. \ref{schematic-wireless-tt}), and
	\item[3)] fully wireless time-frequency transfer (Fig. \ref{schematic}).
\end{itemize}}
Each of the experiments was repeated with \acp{snr} varying from $6$--$36$ dB in $3$-dB increments; at each \ac{snr} level the precision of the time-transfer waveform and beamforming waveforms were recorded. \hlb{An additional sweep of tone separation was included for the fully cabled time-frequency transfer experiment to validate the accuracy trends for varying tone separations relative to the \ac{crlb} while the pre-processing \ac{snr} was held at $30$\,dB.} 
\begin{figure}
	\centering
	\includegraphics[width=\columnwidth]{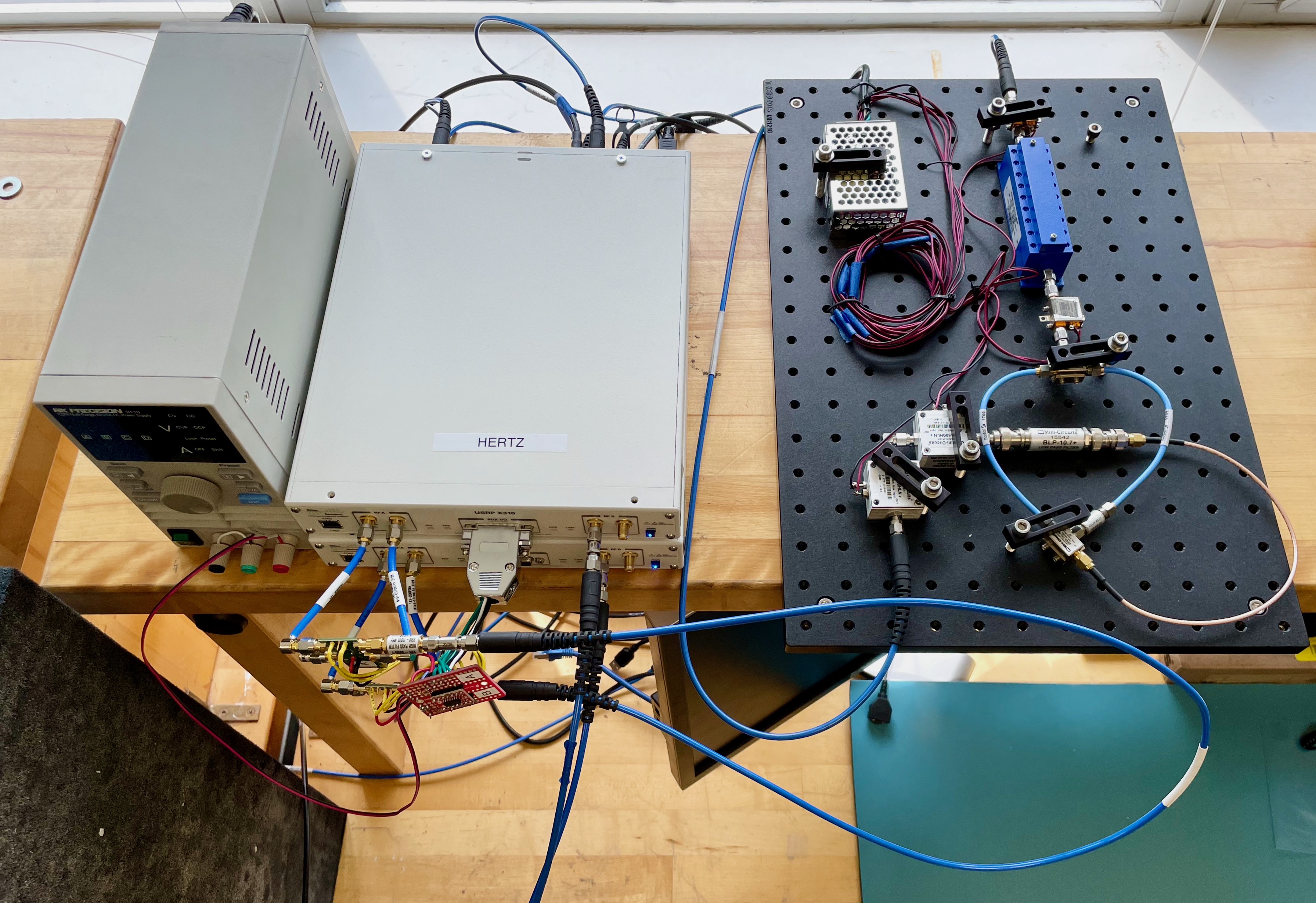}
	\caption{Detail image of the \acp{sdr} in the fully-cabled configuration (left) with the two-tone self-mixing frequency locking circuit (right).}
	\label{detail-setup}
\end{figure}
\begin{figure*}
	\centering
	\includegraphics[width=\textwidth]{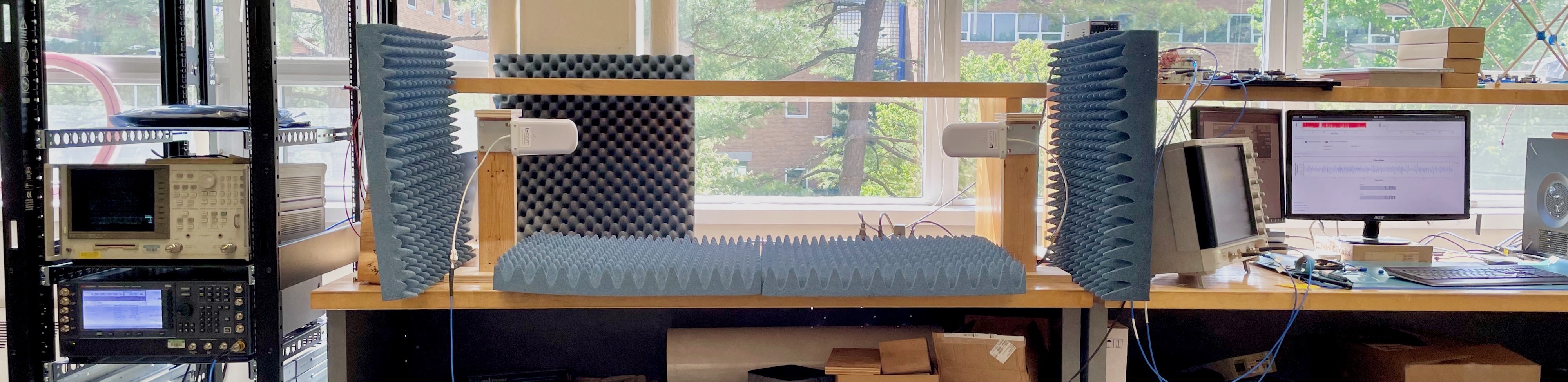}
	\caption{\hlb{Wireless configuration experimental Setup. Signal generator (left) used for the primary \ac{sdr}'s (\ac{sdr} 0) frequency reference for all experiments and two-tone generation in the fully-wireless time-frequency transfer experiment. Time and frequency transfer antennas (center), oscilloscope, \acp{sdr}, and control computer (right).}}
	\label{experimental-setup}
\end{figure*}

\hlb{System schematics for each of the experiments are shown in Figs. \ref{schematic-cabled}--\ref{schematic} and the experimental setups for the wired and wireless configurations are pictured in Figs. \ref{detail-setup}--\ref{experimental-setup}, respectively. A summary of the experimental parameters is provided in Table \ref{tab:experiment-parameters}.} The \acp{sdr} used in these experiments were Ettus Research \ac{usrp} X310's each equipped with two UBX-160 daughterboards which provided \SI{160}{\mega\hertz} of instantaneous analog bandwidth; the X310's were run with a base clock of \SI{200}{\mega\hertz} and a digital sampling rate of \SI{200}{\mega Sa/\second}. To provide high isolation between the transmit and receive paths, two \ac{adi} HMC427A control transfer switches were used and controlled using the \ac{gpio} pins on the \acp{sdr}. Finally, each \ac{sdr} used a bandpass filter to separate the \SI{4.3}{\giga\hertz} frequency-transfer tones from the \SI{5.8}{\giga\hertz} time transfer tones, reducing distortion of the received signals. A \acf{pps} cable was connected between \acp{sdr} for a coarse initial time alignment and is only used once on initialization; this aligned the systems to within several clock ticks which is required to align the finite transmit and receive windows close enough that the time synchronization pulses transmitted would arrive within the receive window. This coarse time alignment could also be achieved fully wirelessly by first starting at a low sampling rate and using low bandwidth waveforms with either continuously streaming receivers or large receive time windows to obtain a coarse inter-\ac{sdr} time offset while accommodating processing power of the host computer; shorter receive windows could be used with progressively higher sample rates to refine the time delay estimate until the full bandwidth of the device is realized, if needed. Other more conventional approaches may also be used such as \ac{gnss} \ac{pps} synchronization, or adjunct \acf{uwb} transmitters if the application permits. For the fully cabled time-frequency transfer and cabled frequency syntonization experiments, the \SI{10}{\mega\hertz} reference output of \ac{sdr} 0 was connected directly to the reference input of \ac{sdr} 1, and for the fully cabled experiment the time transfer was performed over a 3\,ft coaxial cable with a $30$-dB attenuator, shown in Fig. \ref{detail-setup}. During the wireless experiments, two L-Com 8 dBi \SI{2.3}{}--\SI{6.5}{\giga\hertz} log-periodic antennas were placed \SI{90}{\centi\meter} apart (radome to radome) to perform the time and frequency transfer, shown in Fig. \ref{experimental-setup}. 

A Keysight PSG E8267D vector signal generator was used to generate the \SI{10}{\mega\hertz} frequency reference for \ac{sdr} 0 for all experiments, and to generate the two-tone waveform with a \SI{4.295}{\giga\hertz} carrier tone and a single \SI{4.305}{\giga\hertz} sideband used by the self-mixing frequency locking circuit (Fig. \ref{detail-setup}, \cite[Section II]{mghabghab2020frequency}) to generate the \SI{10}{\mega\hertz} reference for \ac{sdr} 1 in the fully wireless experiment; for all other experiments \ac{sdr}~1 was locked to the \SI{10}{\mega\hertz} reference output of \ac{sdr} 0.  In all experiments a Keysight DSOS804A \SI{20}{\giga Sa/\second} oscilloscope configured with an \SI{8.4}{\giga\hertz} analog bandwidth was used to capture the two beamforming waveforms for pulse alignment estimation. 

To control and process the data from the \acp{sdr}, each \ac{sdr} was connected to a desktop computer using \SI{10}{\giga bit} Ethernet. The control and processing computer consisted of a \SI{2.3}{\giga\hertz} Intel i5-2500T processor with \SI{32}{\giga B} of \SI{1333}{\mega\hertz} DDR3 memory running Ubuntu 20.04. GNU Radio 3.9 and the Ettus \acf{uhd} 4.1 were used to interface with the \acp{sdr} and process the data in real-time.  To achieve the full \SI{200}{\mega Sa/\second} performance as well as ensure the most accurate timing, the \acp{sdr} were programmed for bursty operation using timed transmissions/receptions which allowed scheduling of messages to be transmitted and received down to a single local clock tick on each platform.  However, due to API limitations at the time of implementation, an in-place local clock update operation was not supported so local time offsets were stored on the control computer and manually added to the scheduled transmit times to ensure clocks were aligned in software. The clocks were aligned using the two-way time synchronization exchange as described in Section \ref{two-way-time-syncrhonization} with a \ac{pri} of \SI{50}{\milli\second}, pulse duration of \SI{10}{\micro\second}, and two-tone bandwidth of \SI{40}{\mega\hertz}. Each time delay was estimated using a single waveform pulse. A finite rise and fall time of \SI{50}{\nano\second} was applied to the time transfer waveform envelope to generate a waveform which could more realistically be generated by the device with finite switching time, and thus generate a signal which is closer to the ideal signal used for matched filtering. Furthermore, by spreading the rising edge of the envelope across several samples, the exact time of arrival between sample bins can be more easily deduced due to the addition of waveform amplitude modulation.

To determine the accuracy of the secondary beamforming channel with high precision, \SI{50}{\mega\hertz} \SI{1}{\micro\second} two-tone waveforms with a \SI{1.2}{\giga\hertz} carrier frequency were used. \hlb{A two-tone waveform was used for the beamforming signal rather than a typical communications or radar waveform because, as described in Section \ref{accuracy-bounds}, the two-tone waveform provides the optimal accuracy in measuring the timing accuracy of the beamformed signals.} A \SI{1.2}{\giga\hertz} carrier was chosen for its minimal phase noise on the UBX-160 daughterboards. The signals were digitized by the oscilloscope and saved to disk and digitally downconverted and cross-correlated in post-processing using Python to determine their inter-arrival time difference. The standard deviation was then computed for each \ac{snr};
the long-term bias trends were also measured for the maximum \ac{snr} case of 36 dB.
The time-transfer stability was measured using the standard deviation of the self-reported timing corrections based on the two-way time transfer process.

To perform the \ac{snr} control, the transmit gain was first increased until reaching a gain of $30$ dB ($\sim$$+15$ dBm) at which point the receive gain was increased to reach an estimated \ac{snr} of $36$ dB.  To estimate the \ac{snr}, a simple \ac{rms} power method was used. Because both the two-tone and \ac{lfm} waveforms are constant-amplitude pulses, the \ac{rms} signal power could be determined directly from the measured signal envelope in a $50$-$\Omega$ system by
\begin{equation}
	P_{\mathrm{s}} = \sqrt{\frac{1}{N}\sum_{n=1}^{N}\frac{\left|r_{\mathrm{s}}[n]\right|^2}{50}}
\end{equation}
where $N$ is the number of samples received and $r_{\mathrm{s}}$ is the received pulse samples. 
The noise power was similarly estimated from the received signal envelope when there was no transmission occurring by
\begin{equation}
	P_{\mathrm{n}} = \sqrt{\frac{1}{N}\sum_{n=1}^{N}\frac{\left|r_{\mathrm{n}}[n]\right|^2}{50}}
\end{equation}
where $r_{\mathrm{n}}$ is the received signal when there was no transmission was occurring; an equal number of noise samples were used to calculate the noise power.  Finally, the \ac{snr} was estimated by
\begin{equation}
	\mathrm{\ac{snr}}=10\log_{10}{\left[\left(\frac{P_\mathrm{s}}{P_\mathrm{n}}\right)^2\right]}.
\end{equation}

\subsection{Experimental Results}
\label{experimental-results}

\begin{figure}
	\centering
	\includegraphics{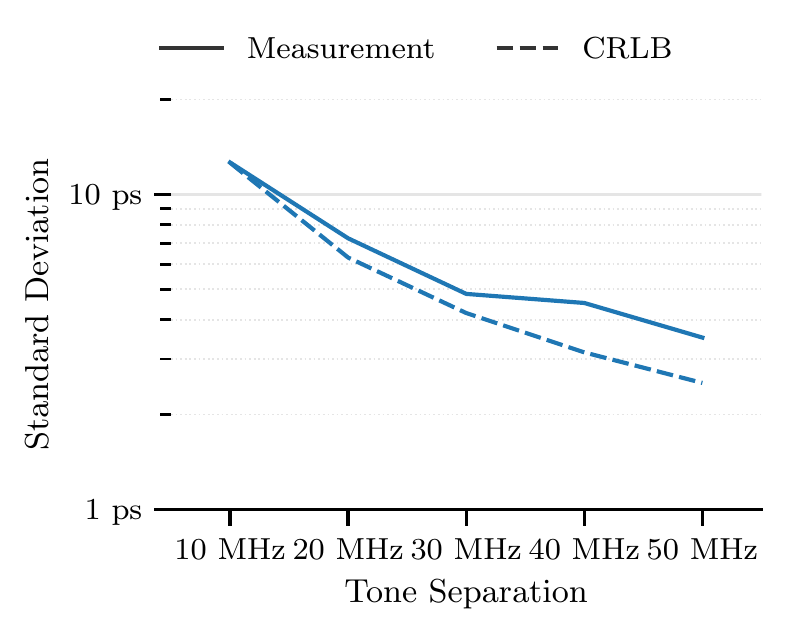}
	\caption{\hlb{Accuracy vs. tone separation relative to the \ac{crlb} for the pulsed two-tone time transfer process when fully cabled.  The tone separation was varied from \SI{10}{}--\SI{50}{\mega\hertz} and measurements were collected at a pre-processing \ac{snr} of $30$\,dB at a carrier of $f_{0\mathrm{t}}=\SI{5.8}{\giga\hertz}$.}}
	\label{bw-sweep}
\end{figure}

\begin{figure*}
	\centering
	\includegraphics{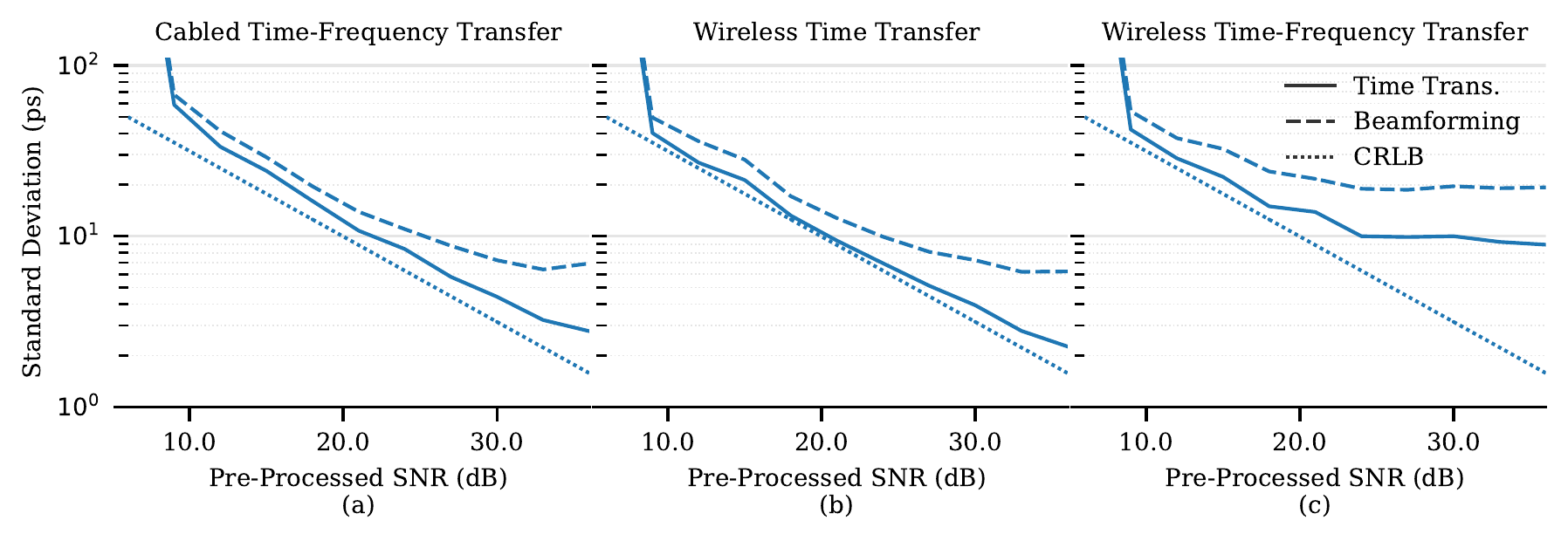}
	\caption{Precision measurements ranging from $6$--$36\,$dB \ac{snr} for each of the (a) cabled time-frequency transfer, (b) wireless time transfer, and (c) wireless time-frequency transfer experiments. The time transfer measurement is the self-reported standard deviation of the time synchronization between \acp{sdr}; the beamforming measurements is the standard deviation of the beamforming pulses set to the oscilloscope; and the \ac{crlb} is the theoretical lower bound computed using \eqref{crlb}--\eqref{msb-two-tone}. The \ac{crlb} is presented for the ``best case'' \ac{snr} within the \ac{snr} estimate uncertainty of $\pm3\,$dB.}
	\label{stats}
\end{figure*}

\begin{figure}
	\centering
	\includegraphics{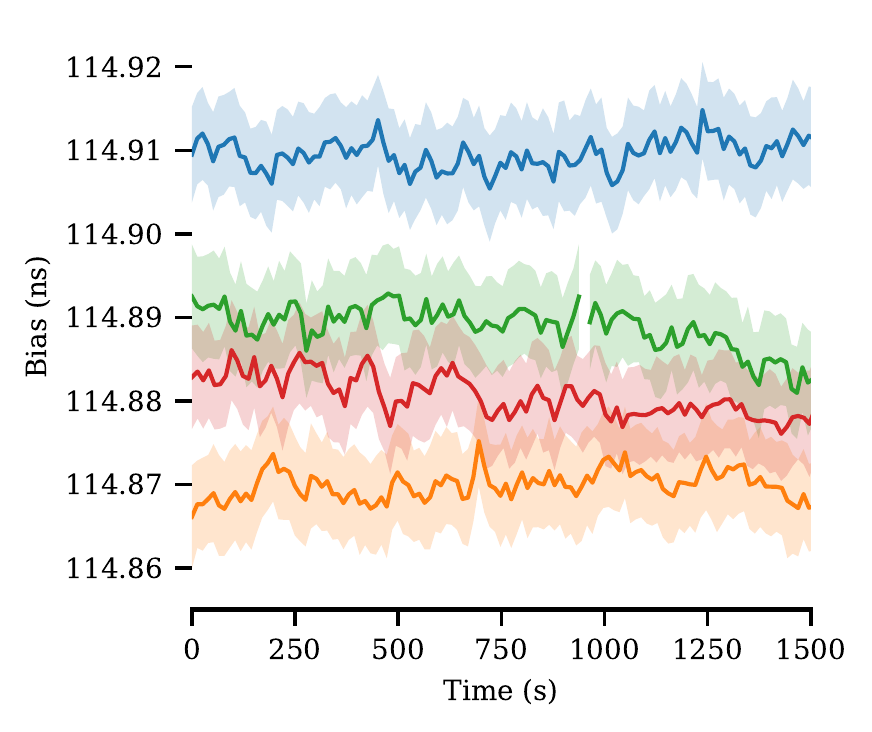}
	\caption{Long-term beamforming bias trends taken across multiple days without constant bias removed. Shaded region depicts one standard deviation. Constant bias of $\sim$\SI{114.89}{\nano\second} due to initial \ac{pps} triggering latency, internal \ac{sdr} delays and external transmission line mismatch; slowly time-varying bias of $<$\SI{100}{\pico\second} believed to be due to a time-varying internal \ac{sdr} clock distribution skew.}
	\label{bias-stats-dynamic}
\end{figure}

During each of these experiments, the synchronization epoch occurred over \SI{50}{\milli\second} intervals and resynchronized every \SI{50}{\milli\second}. Utilizing these parameters, the system precision (standard deviation) and accuracy (standard deviation + bias) were collected.  The precision was measured over a range of \acp{snr} using 1000 beamforming pulses over approximately two minutes, the results of which are shown in \hlb{Figs. \ref{bw-sweep} and \ref{stats} for the bandwidth/tone separation sweep, and \ac{snr} sweeps, respectively. In Fig. \ref{bw-sweep} the measured time transfer accuracy is shown in a solid line while the \ac{crlb} is shown as a dashed line; for both the two-tone and \ac{lfm}, the measured data follows the trend of the \ac{crlb}, however the two-tone approaches more closely to the \ac{crlb}. The time transfer accuracy is denoted in Fig. \ref{stats} in solid blue while the beamforming accuracy, computed from the cross-correlated oscilloscope samples is show as a dashed blue line.} For moderate to high \acp{snr} of $>$15 dB, the beamforming accuracy is typically $\sim$\SI{3}{\pico\second} higher than the time alignment accuracy. Furthermore, it is noted that the experiment using the fully-wireless two-tone time-frequency transfer circuit, summarized in Fig. \ref{stats}\;(c), imposes a lower bound on the time transfer precision of $\sim$\SI{10}{\pico\second} past 15-dB \ac{snr} due to an increased clock phase noise from the frequency transfer circuit. However, the wireless time transfer technique alone using cabled frequency syntonization closely follows that of the fully cabled time-frequency transfer case demonstrating the efficacy of the technique over wireless links.

The long-term bias between the information beamforming channels on each \ac{sdr} was also taken over multiple \SI{1500}{\second} periods across multiple days to demonstrate the typical biases experienced, the results of which are shown in Fig. \ref{bias-stats-dynamic}. The biases consist of a static offset and a small slowly time-varying bias. Factors which contribute to the initial bias are initial \ac{pps} triggering latency, internal device delays, constant inter-channel timing skew, and external transmission line length mismatch between the \ac{sdr} and oscilloscope. This static bias can be calibrated out by measuring the average inter-channel bias over a short period such that the minimal time-varying drift occurs, then switching the channels which the cables are connected to on the \ac{sdr} and repeating the process, then averaging the results of the two measurements to remove any variation in cable and adapter length and oscilloscope input circuitry. The small slowly time-varying bias typically varies by $<$\SI{100}{\pico\second} over long durations (hours to days) which is believed to be due to a small time-varying inter-channel timing skew internal to each \ac{sdr} which causes the sampled signals on each channel, and each data converter within each channel, to be sampled with a slight, relative time-varying skew; these small timing skews are cumulative and can sum to cause the $\sim$\SI{100}{\pico\second} skews observed. Because of this, the long-term accuracy is limited by the inter-device timing skews specific to the device used in this experiment.

\subsection{Discussion}

\begin{table*}
	\caption{Comparison of Current Sub-Nanosecond Microwave and Millimeter-Wave Wireless Time Transfer Technique Stability}
	\label{tab:time-transfer-comparision}
	\begin{center}
  	\begin{tabularx}{\textwidth}{YYYYYYY}
	\toprule[1pt]
	Reference & Waveform* & \hlb{SNR (dB)} & Carrier Frequency (GHz) & Bandwidth (MHz) & Standard Deviation (ps) & Figure of Merit$^\dagger$ (Lower is better) \\
	\midrule
	\midrule
	\cite{alemdar2021rfclock} & 802.15.4\;\ac{uwb}\;(D) & \hlb{NA} & $3.5$--$6.5$ & $\sim$$900$ & $477$ & $\sim$$429,300$ \\
	\midrule
	\cite{seijo2020enhanced} & 802.11n (D) & \hlb{NA} & $2.412$ & $20$ & $\sim$$650$ & $\sim$$13,000$ \\
	\midrule
	\cite{roehr2007method} & \ac{lfm} & \hlb{NA} & $5.725$ & $150$ & $66$ & $9,900$ \\
	\midrule
	\cite{pooler2018precise} & \ac{lfm} & \hlb{NA} & $3.120$ & $40$ & $<100$ & $<4,000$ \\
	\midrule
	\cite{gilligan2020white} & White Rabbit (D) & \hlb{NA} & $72.0$--$75.0$ & $1,600$ & $<2.0$ & $<3,200$\\
	\midrule
	\cite{prager2020wireless} & \ac{lfm} & \hlb{$31.2$} & $1.0$ & $50$ & $11.3$ & $565$ \\
	\midrule
	\hlb{This Work} & \hlb{Two-Tone} & \hlb{$30.0$} & \hlb{$5.8$} & \hlb{$40$} & \hlb{$3.94$} & \hlb{$157.6$} \\
	\midrule
	\textbf{This Work} & \textbf{Two-Tone} & \hlb{\textbf{36.0}} & \textbf{5.8} & \textbf{40} & \textbf{2.26} & \textbf{90.4} \\
	\bottomrule[1pt]
	\end{tabularx}
	\end{center}
	\footnotesize{*(D) = digitally modulated waveform\\$^\dagger$\Ac{fom} = bandwidth (MHz) $\times$ standard deviation (ps)\\\hlb{NA: Not available at time of publication}}
\end{table*}

As discussed in Section \ref{experimental-results}, the total accuracy is limited by the slowly time-varying inter-channel hardware bias; if left uncorrected this could reduce the overall system beamforming bandwidth. However, this could be corrected in a similar way to the inter-system time transfer process, by periodically performing an intra-system time transfer operation to remove the self-bias between channels.  
It should also be noted that an online optimization system could be utilized to determine the optimal times to perform clock updates based on overall system drift characterization.  In this paper, we demonstrate a constant periodic synchronization, however, if the system is static with well syntonized clocks, it may be beneficial to reduce the resynchronization frequency to avoid jitter in the timing.  This could be implemented by periodically checking for inter-device timing skew and allowing a time-offset correction only when it is outside the tolerable limits.  This would reduce the overall beamforming jitter in cases where the inter-\ac{sdr} bias is smaller than the precision of the time transfer link, e.g., in low-\ac{snr} environments or when very high-quality oscillators are used, while still maintaining a high level of timing coherence between the systems.
In addition, as described in the \ac{crlb} given by \eqref{crlb} and \eqref{tbp}, and verified experimentally, the accuracy of the time delay estimate is improved by increasing \ac{snr}; the \ac{snr} of the system may be increased by means of increasing transmit power or reducing system noise, however, these gains become significantly more difficult at higher \ac{snr}.  

Finally, a comparison of the results in this work to other similar microwave and millimeter-wave wireless time transfer systems is shown in Table \ref{tab:time-transfer-comparision}.  A \ac{fom} to rank the achieved system time transfer precision versus the occupied signal bandwidth is defined as the product of the signal bandwidth in MHz and the time transfer standard deviation in picoseconds; thus, a lower \ac{fom} indicates better timing performance with a lower signal bandwidth. 
\hlb{We chose this \ac{fom} because it captures the controllable aspect of the waveform (the bandwidth) along with the performance (the time synchronization standard deviation). While an alternative metric may include the \ac{snr}, very few other works in the literature report this value.}
The time transfer approach demonstrated in this work yielded a \ac{fom} of 90.4 \hlb{with a \ac{snr} of $36$\,dB, and 157.6 with an \ac{snr} of $30$\,dB. This \ac{snr} value is comparable to one other work that reported \ac{snr}, \cite{prager2020wireless}, which had an \ac{snr} of 31.2 dB but which achieved a \ac{fom} of 565, far higher than that reported in this work.
It is important to note that all other waveforms used in time transfer works in the literature were filled bandwidth waveforms, such as \acp{lfm}\cite{roehr2007method,pooler2018precise,prager2020wireless} and digitally encoded waveforms \cite{alemdar2021rfclock,seijo2020enhanced,gilligan2020white}.} In contrast, in our approach the waveform uses only two tones at the ends of the bandwidth; thus, the bandwidth between the tones can be left unused to reduce bandwidth requirements on the system by implementing only two narrow-band signal generators to produce the two-tone pulses, or it can be used for other wireless operations, such as coarse PPS or inter-node communications; this is not possible with other filled-bandwidth waveforms. 
\hlb{While the performance to spectral efficiency of the two-tone time transfer method is significantly greater than the other conventional filled bandwidth techniques listed, it should be noted that the accuracy of some of the other works were evaluated outside of laboratory environments which may reduce accuracy due to multipath and uncontrollable environmental dynamics.}
Using this method it is clear that the two-tone time transfer waveform provides a bandwidth-efficient technique for achieving high accuracy wireless time transfer in distributed wireless systems.

\section{Conclusion}
\hlb{In this paper, we demonstrated the first fully wireless time transfer system capable of synchronizing time between two systems to a precision of \SI{2.26}{\pico\second} over a $36$-dB \ac{snr} wireless link using a novel single pulse two-tone time delay estimation technique which achieves the highest known theoretical accuracy for a given signal bandwidth.} This shows a significant step towards improving the overall system accuracy towards sub-picosecond timing alignment using \ac{rf} systems enabling high accuracy coordination in wireless distributed arrays for high bandwidth distributed antenna arrays.

\bibliographystyle{IEEEtran}
\bibliography{IEEEabrv, wireless_time_frequency_synchronization.bib}

\begin{IEEEbiography}[{\includegraphics[width=1in,height=1.25in,clip,keepaspectratio]{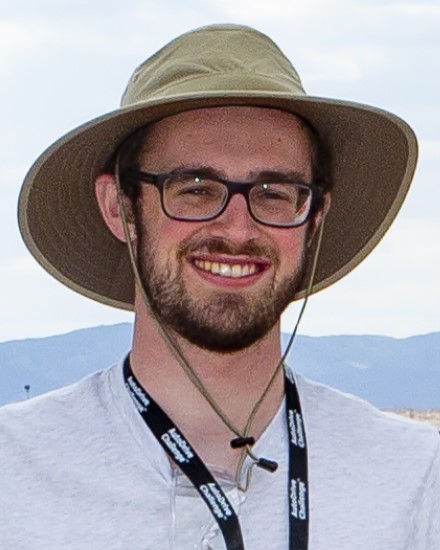}}]{Jason M. Merlo} (Graduate Student Member, IEEE) received the B.S. degree in computer engineering from Michigan State University, East Lansing, MI, USA in 2018, where he is currently pursuing the Ph.D. degree in electrical engineering. 

From 2017-2021 he was project manager and electrical systems team lead of the Michigan State University AutoDrive Challenge team. His current research interests include distributed radar and wireless system synchronization, interferometric arrays, synthetic aperture radar, joint radar-communications, and automotive/automated vehicle radar applications.
\end{IEEEbiography}
\vskip 0pt plus -1fil

\begin{IEEEbiography}[{\includegraphics[width=1in,height=1.25in,clip,keepaspectratio]{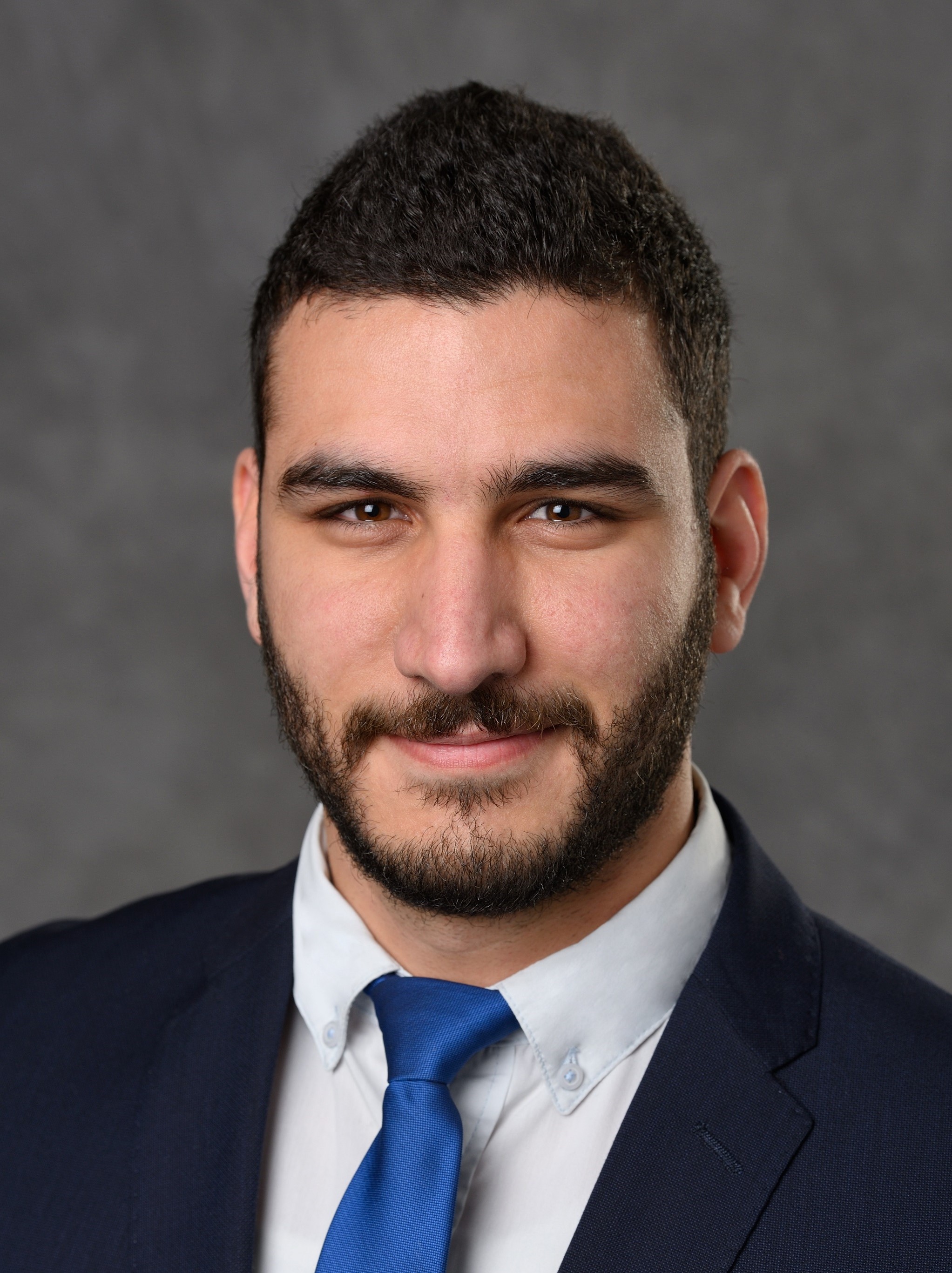}}]{Serge R. Mghabghab} (Member, IEEE) received the B.E. degree in electrical engineering from Notre Dame University Louaize, Zouk Mosbeh, Lebanon in 2014 and the M.E. degree in electrical and computer engineering from American University of Beirut, Beirut, Lebanon in 2017. In 2022 he received the Ph.D. degree in electrical engineering at Michigan State University, East Lansing, MI, USA.

In 2022 he joined MathWorks in Natick, MA, USA.
\end{IEEEbiography}
\vskip 0pt plus -1fil

\begin{IEEEbiography}[{\includegraphics[width=1in,height=1.25in,clip,keepaspectratio]{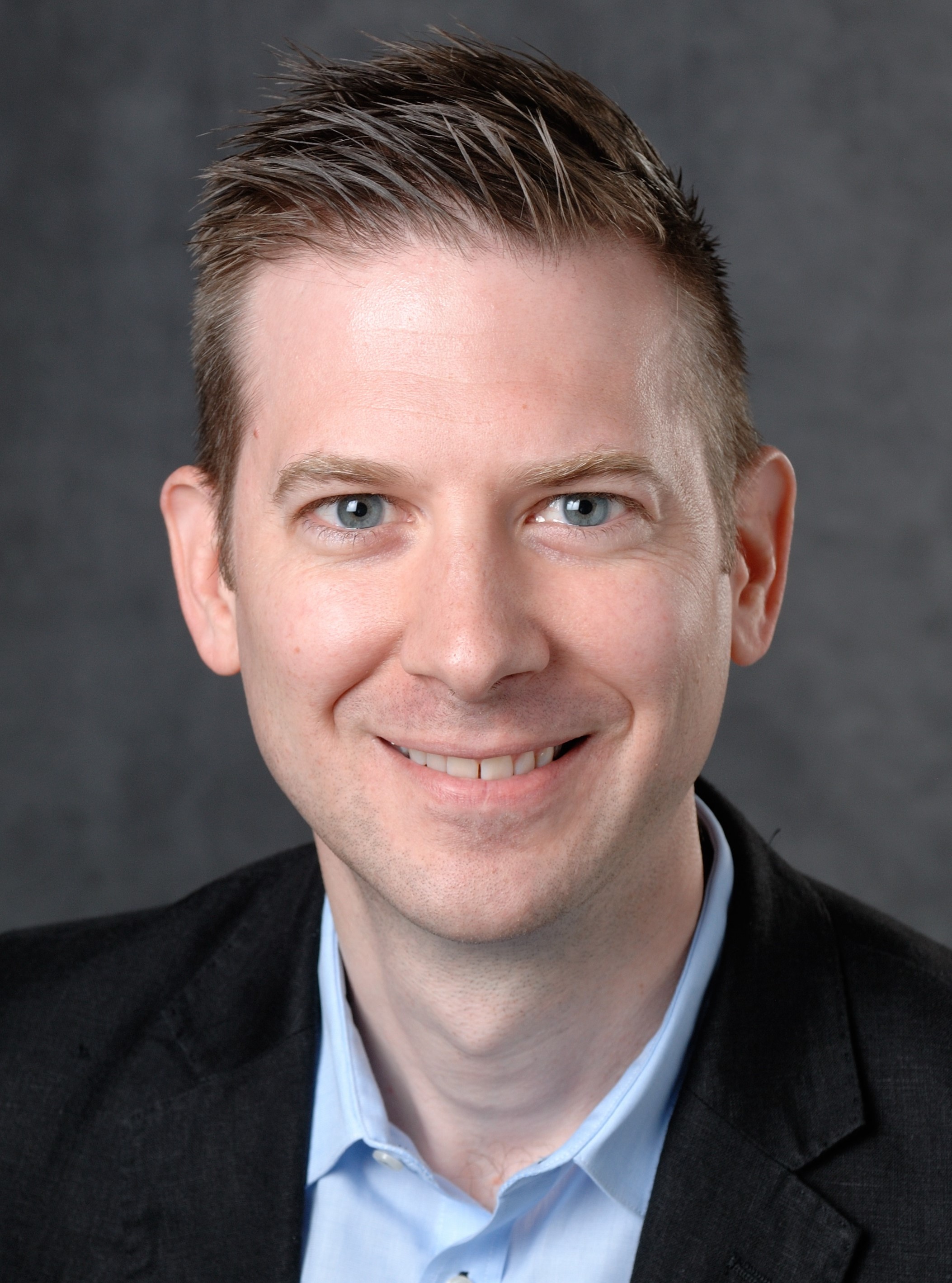}}]{Jeffrey A. Nanzer} (Senior Member, IEEE) received the B.S. degrees in electrical engineering and in computer engineering from Michigan State University, East Lansing, MI, USA, in 2003, and the M.S. and Ph.D. degrees in electrical engineering from The University of Texas at Austin, Austin, TX, USA, in 2005 and 2008, respectively.

From 2008 to 2009 he was with the University of Texas Applied Research Laboratories in Austin, Texas as a Post-Doctoral Fellow designing electrically small HF antennas and communications systems. From 2009 to 2016 he was with the Johns Hopkins University Applied Physics Laboratory where he created and led the Advanced Microwave and Millimeter-Wave Technology Section. In 2016 he joined the Department of Electrical and Computer Engineering at Michigan State University where he held the Dennis P. Nyquist Assistant Professorship from 2016 through 2021. He is currently an Associate Professor. He directs the Electromagnetics Laboratory, which consists of the Antenna Laboratory, the Radar Laboratory, and the Wireless Laboratory. He has published more than 200 refereed journal and conference papers, two book chapters, and the book  Microwave and Millimeter-Wave Remote Sensing for Security Applications (Artech House, 2012). His research interests are in the areas of distributed phased arrays, dynamic antenna arrays, millimeter-wave imaging, remote sensing, millimeter-wave photonics, and electromagnetics.

Dr. Nanzer is a Distinguished Microwave Lecturer for the IEEE Microwave Theory and Techniques Society (Tatsuo Itoh Class of 2022-2024). He was a Guest Editor of the Special Issue on Special Issue on Radar and Microwave Sensor Systems in the IEEE Microwave and Wireless Components Letters in 2022. He is a member of the IEEE Antennas and Propagation Society Education Committee and the USNC/URSI Commission B, was a founding member and the First Treasurer of the IEEE APS/MTT-S Central Texas Chapter, served as the Vice Chair for the IEEE Antenna Standards
Committee from 2013 to 2015, and served as the Chair of the Microwave Systems
Technical Committee (MTT-16), IEEE Microwave Theory and Techniques
Society from 2016 to 2018.
He was a recipient of the Google Research Scholar Award in 2022, the IEEE MTT-S Outstanding
Young Engineer Award in 2019, the DARPA Directors Fellowship in 2019, the National Science Foundation (NSF) CAREER Award in 2018, the DARPA Young
Faculty Award in 2017, and the JHU/APL Outstanding Professional Book
Award in 2012.  He is currently an Associate Editor of the \textsc{IEEE Transactions on Antennas and Propagation}.
\end{IEEEbiography}

\end{document}